\documentclass[prd,aps,preprint,nofootinbib,epsfig,floats]{revtex4}

\usepackage{graphicx}
\usepackage{times}
\usepackage{epsfig}
\usepackage{amsmath}
\usepackage{amsfonts}
\usepackage{amssymb}
\usepackage{url}
\usepackage{hyperref}
\usepackage{subfigure}
\newcommand{\be}{\begin{equation}}
\newcommand{\ee}{\end{equation}}
\newcommand{\bea}{\begin{eqnarray}}
\newcommand{\eea}{\end{eqnarray}}

\begin{document}

\pagestyle{plain}

\title{Supersymmetry Breaking by Type II Seesaw Assisted Anomaly
Mediation }

\author{R. N. Mohapatra\footnote{
  e-mail: rmohapat@physics.umd.edu},
       }

\affiliation{Department of Physics,
 University of Maryland, College Park, MD 20742, USA}
\author{Nobuchika Okada\footnote{
  e-mail: okadan@post.kek.jp},}
\affiliation{ Theory Division, KEK, 1-1 Oho, Tsukuba, 305-0801,
Japan}

\author{Hai-Bo Yu\footnote{
  e-mail: haiboy@uci.edu}}
\affiliation{Department of Physics and Astronomy,
 University of California, Irvine, CA 92697, USA}

\date{November, 2007}

\preprint{\vbox{\hbox{UMD-PP-07-014}}}
\preprint{\vbox{\hbox{KEK-TH-1197}}}
\preprint{\vbox{\hbox{UCI-TR-2007-46}}}

\begin{abstract}
Anomaly mediated supersymmetry breaking (AMSB), when implemented in
MSSM is known to suffer from the problem of negative slepton mass
squared leading to breakdown of electric charge conservation. We
show however that when MSSM is extended to explain small neutrino
masses by including a pair of superheavy Higgs triplet superfields
(the type II seesaw mechanism), the slepton masses can be deflected
from the pure AMSB trajectory and become positive. In a simple model
we present in this paper, the seesaw scale is about
$10^{13}-10^{14}{\rm GeV}$. Gauge coupling unification can be
maintained by embedding the triplet to $SU(5)$ {\bf 15}-multiplet.
In this scenario, bino is the LSP and its mass is nearly degenerate
with NLSP slepton when the triplet mass is right around the seesaw
scale.

\end{abstract}

\maketitle

\section{Introduction \label{sec1}}
Supersymmetry (SUSY) is considered to be a prime candidate for TeV
scale physics since it resolves several conceptual issues of the
standard model (SM) such as (i) radiative stability of the large
hierarchy between Planck and weak scale; and (ii) electroweak
symmetry breaking. With additional assumptions, it develops other
appealing features: for instance, if R-parity symmetry is assumed,
it can provide a candidate for the dark matter of the universe and
if no or specific new physics is assumed, it can lead to the
unification of gauge couplings at a very high scale.

Since there is no trace of supersymmetry in current observations, it
must be a broken symmetry and the question arises as to the origin
of this breaking. While at the phenomenological level, it is
sufficient to assume soft breaking terms to implement this, low
energy observations in the domain of flavor changing neutral
currents (FCNC) imply strong constraints on it i.e. the sparticle
masses must be flavor degenerate. It is therefore reasonable to
require that any mechanism for SUSY breaking must lead to such
flavor degeneracy for slepton and squark masses. Indeed there exist
at least two well known scenarios where this happens: gauge mediated
SUSY breaking (GMSB) \cite{GMSB1} \cite{GMSB2} and anomaly mediated
SUSY breaking (AMSB) \cite{AMSB1} \cite{AMSB2}. In both these cases
in simplest examples, the FCNC effects are dynamically suppressed.
Both involve unknown physics in the hidden sector which breaks
supersymmetry and this SUSY breaking information is transmitted to
the visible sector via certain messengers. In GMSB scenario, the
messenger sector generically involves new particles and forces
whereas in the AMSB scenario, SUSY breaking is transmitted via the
conformal breaking induced by radiative corrections in
supersymmetric field theories. They however differ in the way the
SUSY breaking manifests in the low energy sector: in GMSB (as in
gravity induced minimal SUGRA models), the detailed pattern of
sparticle masses depend on ultraviolet physics i.e. physics at mass
scales much higher than the SUSY breaking scale whereas AMSB models
have the advantage that this pattern depends only on the low scale
physics. They are therefore easier to test experimentally given a
particular low scale theory.

It however turns out that AMSB models despite their elegance and
predictive power suffer from a fatal problem when the low scale
theory is assumed to be the MSSM i.e. they predict the slepton mass
squared to be negative and hence lead to a vacuum state that breaks
electric charge conservation (called tachyonic slepton problem
henceforth). This is of course unacceptable and this problem needs
to be solved if AMSB models have to be viable. There are many
attempts to solve this problem by taking into account additional
positive contribution to the slepton mass squared \cite{dAMSB1}
\cite{amsb4} \cite{dAMSB2} \cite{amsb3}.

An important thing to realize at this point is that MSSM is not a
complete theory of low energy particle physics and needs
extension to explain the small neutrino masses observed in
experiments. The relevant question then is whether MSSM extended
to include new physics that explains small neutrino masses will
cure the tachyonic slepton mass pathology of AMSB.

There are two simple extensions of MSSM which provide natural
explanation of small neutrino masses: the two types of seesaw
mechanisms i.e. type I \cite{seesaw} and type II \cite{type2}. In
the first case, a reasonable procedure is to extend the gauge
symmetry of MSSM to $SU(2)_L\times SU(2)_R\times U(1)_{B-L}\times
SU(3)_c$ which automatically introduces three right-handed neutrinos
into the theory as well as new couplings involving the leptons which
one could imagine as affecting the slepton masses. In most
discussions of seesaw mechanism, it is commonly assumed that the
seesaw scale is very high ($\geq 10^{13}$ GeV or so); so one would
expect the associated new physics interactions to decouple. Such a
generic scenario will not solve the tachyonic slepton problem.
However, it has recently been pointed out \cite{mss} that there
exists a class of minimal SUSY left-right symmetric models with high
scale seesaw where left-handed weak iso-triplets with B-L=+2 and
doubly charged Higgs fields with B-L=+2 coupling to right-handed
leptons have naturally weak scale mass because of higher symmetries
of superpotential. Their couplings to leptons contribute to the
slepton mass squared and can solve the tachyonic slepton mass
problem \cite{mss}.

The present paper focuses on an alternative approach which uses type
II seesaw mechanism for neutrino masses and to see how it affects
the slepton masses. An advantage of this over the type I approach is
that it does not involve extending the gauge symmetry but requires
adding a pair of $Y=\pm 1$ $SU(2)_L$ triplet Higgs fields to MSSM.
The $SU(2)_L$-triplets have mass close to $10^{13}$ GeV which is
required to implement type II seesaw for small neutrino masses. We
further assume that the triplet masses arise from the vacuum
expectation value (VEV) of a light singlet field with a high VEV. We
then show that in AMSB scenario, the F-component of the singlet
field acquires an induced VEV, leading to new set of SUSY breaking
effects. These effects are gauge mediated contributions to sparticle
masses in addition to the usual AMSB contributions. We find that
these contributions solve the tachyonic slepton mass problem. Thus
type II seesaw in addition to solving neutrino mass problem also
solves the problem of SUSY breaking by AMSB \footnote{We note that
pure gauge mediation in the presence of type II seesaw has been
considered recently \cite{rossi}; our model is different since AMSB
effects play a significant role in the final predictions.}. Of
course in this case one needs to assume R-parity symmetry to obtain
stable dark matter.

This scenario makes prediction for the sparticles which are
different from other scenarios. In particular, we find that the bino
and sleptons are nearly degenerate with messenger at the seesaw
scale- a situation which is particularly advantageous for
understanding the dark matter abundance in the universe
\cite{Arnowitt:2007nt}. We also show that the model does preserve
the unification of couplings.

The paper is organized as follows. In Section~\ref{sec2},
we explain the scenario of ``deflected anomaly mediation''
which plays a crucial role in our solution to the tachyonic slepton problem.
In Section~\ref{sec3},
we present a simple superpotential for the singlet field and
calculate the deflection parameter. Section~\ref{sec4}
contains the general formulas of sparticle masses
in the deflected anomaly mediation. In Section~\ref{sec5},
we present the minimal model to solve the tachyonic slepton problem
as well as generate light neutrino masses.
Section~\ref{sec6} contains the extended models which preserve
the gauge coupling unification.
We summarize our results in Section~\ref{sec7}.
In the Appendix A, we present the calculation of the lifetime
of SUSY breaking local minimum.

\section{Deflected Anomaly mediation and Messenger sector \label{sec2}}
It is well known that in the absence of additional supersymmetry
breaking, the AMSB contribution to sparticle masses is ultraviolet
insensitive. It has however been proposed that presence of
additional SUSY breaking effects could deflect the sparticle masses
from the AMSB trajectory and lead to new predictions for sparticle
spectrum. This has been called ``deflected anomaly mediation''
scenario \cite{dAMSB1} \cite{dAMSB2}.  A key ingredient of this
scenario is the presence of gauge mediated contributions arising
from new interactions in the theory. Typically they involve the
introduction of messengers $\Psi$ and $\overline{\Psi}$ with the
following coupling: \bea
 W = S \overline{\Psi} \Psi .
\label{messenger} \eea Clearly $\overline{\Psi}$ and $\Psi$ are the
messenger chiral superfields
 in a vector-like representation under the SM gauge group,
 and $S$ is the singlet superfield.
It is crucial for the messenger fields to be non-singlets,
 at least, under the $SU(2)_L \times U(1)_Y$ gauge group. In our model, the
the $SU(2)_L$ triplets which enforce the type II seesaw will play
the role of these fields \footnote{In order to implement type II
seesaw in the MSSM, we only need one pair of triplets and it turns
out that one pair of triplets is sufficient to lift slepton masses
and leave bino as the LSP.}. Once the scalar component ($S$) and the
$F$ component ($F_S$) in the singlet chiral superfield develop VEVs,
the scalar lepton obtains new contributions to its mass squared
through the same manner as in the gauge mediation scenario
 \cite{GMSB1} \cite{GMSB2}. In our case, $F_S$ is induced by the hidden sector
conformal compensator SUSY breaking. The effect of non-zero $F_S$ is
to deflect the sparticle masses from the pure AMSB trajectory of the
renormalization group equations, thereby solving the tachyonic
slepton problem.

As just noted an important difference between the deflected AMSB
from GMSB is that the SUSY breaking in the messenger sector
is induced by the anomaly mediation, namely, $F_\phi$,
a non-zero $F$ component of the compensator field,
and $F_\phi$ therefore is the unique source of SUSY breaking in this
scenario. Therefore, we can parameterize the SUSY breaking order
parameter in the messenger sector such as
\bea
 \frac{F_S}{S} =d F_\phi .
\eea
Here, $d$ is the so-called ``deflection parameter''
 which characterizes how much the sparticle masses are
 deflected from the pure AMSB results.
Theoretical consistency constrains it to be $ |d| < {\cal O}(1)$,
 because $F_S/S$ is not the original SUSY breaking sector.

We consider a simple model which provides
 a sizable deflection parameter $|d|={\cal O}(1)$.
Let us begin with the supergravity Lagrangian for $S$
 in the superconformal framework
 \cite{Cremmer:1982en} \cite{Kugo:cu}
 (supposing SUSY breaking in the hidden sector and
 fine-tuning of the vanishing cosmological constant),
\begin{eqnarray}
{\cal L} = \int  d^4 \theta \;
\phi^\dagger \phi  \; S^\dagger S
+ \left\{
  \int d^2 \theta \; \phi^3 W(S) + {\rm H.c.}  \right\} ,
\label{basic}
\end{eqnarray}
 where we have assumed the canonical Kahler potential
 (in the superconformal framework),
 $W$ is the superpotential (except for Eq.~(\ref{messenger})),
 and $\phi=1 +\theta^2 F_\phi$ is the compensating multiplet
 with the unique SUSY breaking source $F_\phi$,
 taken to be real and positive through $U(1)_R$ phase rotation.

The scalar potential can be read off as
\begin{eqnarray}
V= |F_S|^2 - S^\dagger S  |F_\phi|^2
   -3 F_\phi W -3 F_\phi^\dagger W^\dagger
\label{potential}
\end{eqnarray}
with the auxiliary field given by
\begin{eqnarray}
  F_S = - \left( S F_\phi  + W_S^\dagger  \right) \; ,
\label{FSterm}
\end{eqnarray}
where $W_S$ stands for $\partial W/\partial S$.

Using the stationary condition $\partial V/\partial S=0$
 and Eq.~(\ref{FSterm}), we can describe the deflection parameter
 in the simple form,
\begin{eqnarray}
 \frac{F_S}{S} = d F_\phi
 = - 2 \frac{W_S}{S W_{SS}} \; F_\phi ,
 \label{FSterm2}
\end{eqnarray}
where $W_{SS}$ stands for $\partial^2 W/\partial S^2$.
This is a useful formula, from which we can understand
 that $S$ should be light in the SUSY limit
 in order to obtain a sizable deflection parameter
 $|d| = {\cal O}(1)$
 because the SUSY mass term ($W_{SS}$) appears
 in the denominator.

\section{Singlet superpotential and deflection parameter \label{sec3}}
As a simple model, let us consider a superpotential
\begin{eqnarray}
W = - m S^2 + \frac{S^4}{M},
 \label{superpotential}
\end{eqnarray}
where $m$ and $M$ are mass parameters, and
 we assume them to be real, positive and $m \ll M$ \footnote{We have checked that there
are no large scalar S mass terms induced by loop corrections in the
theory.}. The scalar potential is given by \bea
 V=|S|^2 \left| -2 m+4 \frac{S^2}{M} \right|^2
  + F_\phi \left( m S^2 + \frac{S^4}{M} \right) +{ \rm H.c.}
 \label{scalar-potential}
\eea
Changing a variable as $S^2 =x e^{i \varphi}$
 with real parameters, $x \geq 0$ and $0 \leq \varphi \leq 2 \pi$,
 the scalar potential is rewritten as
\bea
 V(x, \varphi)= 4 x \left(
   m^2 - 4 \frac{m}{M} x \cos(\varphi) +4 \frac{x^2}{M^2}
  \right)
 + 2 F_\phi \left( m  x \cos (\varphi)
 +  \frac{x^2}{M} \cos(2 \varphi)  \right).
\eea
It is easy to check that $\varphi=0$ satisfies
 the stationary condition $\partial V/\partial \varphi =0$,
 and we take $\varphi=0$.
Solving the stationary condition
 $\partial V(x,\varphi=0)/ \partial x =0 $,
 we find
\bea
 x_{\pm} = \frac{M}{24}
  \left(
   8 m -F_\phi \pm \sqrt{D}
 \right) ,
\eea
 where $ D= 16 m^2 -40 F_\phi m +F_\phi^2$.
It is easy to show that $x_{+}$ and $x_{-}$
 corresponding to local minimum and maximum of
 the potential, respectively.
For a fixed $F_\phi$, the potential minimum exists
 if $ D >0 $, in other words,
\bea
 m > \frac{5+2 \sqrt{6}}{4} F_\phi  .
\eea

 From Eq.~(\ref{FSterm2}), the deflection parameter is give by
\bea
  d = \frac{-2 m+4 x_+/M}{m-6x_+/M}
    = \frac{2 (4 m + F_\phi -\sqrt{D})}{3 (4 m -F_\phi+\sqrt{D}) } .
\eea
The deflection parameter reaches its maximum value ($d_{\rm max}$)
 in the limit $ m \rightarrow \frac{5+2 \sqrt{6}}{4} F_\phi$,
 and
\bea
  d_{\rm max}=  \frac{2 (3 + \sqrt{6})}{3 (2 +\sqrt{6})}
 \simeq 0.816 .
\eea Squared masses of two real scalar fields in $S=(x+iy)/\sqrt{2}$
 are found to be
\bea
 m_x^2 &=&  8 \frac{\sqrt{D} x_+}{M},   \nonumber \\
 m_y^2 &=& \frac{2}{3}
  \left(
    24 m F_\phi+(2 m-F_\phi) \sqrt{D}+D^2   \right),
\eea
 which are roughly of order $m^2$.
Through numerical calculation, we find
  $m_x \simeq 0.24 F_\phi$ and $m_y \simeq 6.3 F_\phi$
  for $m$ very close to its minimum value leading to $d=0.81$.

The scalar potential of Eq.~(\ref{scalar-potential}),
 in fact, has a SUSY minimum at $S=0$,
 where the potential energy is zero, and the minimum
 at $x_+$ we have discussed is a local minimum.
In the Appendix A, we estimate the decay rate of the local minimum
to the true SUSY minimum and find it is sufficiently small for
$F_\phi \ll M$.

\section{Sparticle mass spectrum \label{sec4}}
We first give general formulas for sparticle masses
 in the deflected anomaly mediation
 with non-zero deflection parameter $d$.
Following the method developed in Ref.~\cite{Giudice:1997ni}
 (see also Ref.~\cite{dAMSB1}),
 we can extract the sparticle mass formulas
 from the renormalized gauge couplings ($ \alpha_i (\mu, S) $)
 and the supersymmetric wave function renormalization coefficients
 (${Z}_I(\mu,S)$) at the renormalization scale ($\mu$)
 and the messenger scale ($S$).
With $F_S/S=d F_\phi$, the gaugino masses ($M_i$)
 and sfermion masses ($\tilde{m}_I$) are given by
\bea
\frac{M_i}{\alpha_i(\mu)}
&=& \frac{F_\phi}{2} \left(
\frac{\partial}{\partial \mbox{ln} \mu}
- d \frac{\partial}{\partial \mbox{ln}|S|}
 \right) \alpha_i^{-1} (\mu, S) ,  \nonumber \\
\tilde{m}^2_I(\mu) &=&
-\frac{|F_\phi|^2}{4}
 \left(
  \frac{\partial}{\partial \mbox{ln} \mu}
  - d \frac{\partial}{\partial \mbox{ln}|S|}  \right)^2
  \mbox{ln} Z_I(\mu, S) .
\label{softmass}
\eea

For a simple gauge group, the gauge coupling and
 the wave function renormalizations are given by
\bea
\alpha_i^{-1} (\mu, S)
&=&
 \alpha_i^{-1} (\Lambda_{cut})
+ \frac{b_i-N_i}{4 \pi}
  \mbox{ln} \left(
 \frac{S^\dagger S}{\Lambda_{cut}^2} \right)
 + \frac{b_i}{4 \pi} \mbox{ln} \left(
 \frac{\mu^2}{S^\dagger S} \right)  \; ,
\label{gaugino} \\
{Z}_I (\mu, S) &=& \sum_i {Z}_I(\Lambda_{cut} )
\left(\frac{\alpha_i(\Lambda_{cut})}{\alpha_i(S)} \right)^{\frac{2
c_i}{b_i-N_i}} \left(\frac{\alpha_i(S)}{\alpha_i(\mu)}
\right)^{\frac{2 c_i}{b_i}},
 \label{scalar}
\eea
where $\Lambda_{cut}$ is the ultraviolet cutoff,
 $b_i$ are the beta function coefficients for different groups,
 $c_i$ are the quadratic Casimirs,
 $N_i$, the Dynkin indices of the corresponding messenger fields
 (for example, $N_i=1$ for a vector-like pair of
  messengers of a fundamental representation
  under $SU(N)$ gauge group),
 and the sum is taken corresponding to the representation
 of the sparticles under the SM gauge groups.
Substituting them into Eq.~(\ref{softmass}), we obtain
\bea
M_i(\mu ) &=&
 \frac{\alpha_i(\mu)}{4 \pi} F_\phi (b_i+ d N_i)  ,
 \label{gauginomass}  \\
\tilde{m}_I^2 (\mu) &=&
 \sum_i
2 c_i \left(
\frac{\alpha_i(\mu)}{4 \pi} \right)^2
 |F_\phi|^2 \; b_i \; G_i(\mu, S) \; ,
\label{scalarmass}
\end{eqnarray}
where
\begin{eqnarray}
G_i(\mu, S) =
\left( \frac{N_i}{b_i} \xi_i^2 + \frac{N_i^2}{b_i^2} (1-\xi_i^2)
 \right) d^2
+ 2 \frac{N_i}{b_i} d  + 1
\label{G}
\eea
with
\begin{eqnarray}
\xi_i  \equiv  \frac{\alpha_i(S)}{\alpha_i(\mu)}
= \left[ 1+ \frac{b_i}{4 \pi} \alpha_i(\mu) \mbox{ln}
\left(
\frac{S^\dagger S}{\mu^2} \right)
 \right]^{-1}  .
\label{xi}
\end{eqnarray}
In the limit $d \rightarrow 0$,
 the pure AMSB results are recovered
 and Eq.~(\ref{scalarmass}) leads to the mass squared
 negative for an asymptotically non-free gauge theory ($b_i < 0$ ).
This result causes the tachyonic slepton problem
 in the pure AMSB scenario.

After integrating the messengers out, the
 scalar mass squared at the messenger scale is given by
(taking $\xi_i=1$) \bea {\tilde m}_I^2 (S) &=&
 \sum_i
2 c_i \left(
\frac{\alpha_i(S)}{4 \pi} \right)^2
 |F_\phi|^2
\left[N_i d^2 +2 N_i d +b_i
\right] ;
 \label{at-threshold}
\eea
where the first, the second and the third terms in the brackets
 correspond to pure GMSB, mixed GMSB and AMSB, and pure AMSB
 contributions, respectively.
The sign of the second term is proportional to $d$,
 so that the sign of the deflection parameter
 results in different sparticle mass spectrum.
The case $d<0$ has been investigated in Ref.~\cite{dAMSB1}
 and the resultant sparticle mass spectrum at the electroweak scale
 is very unusual and colored sparticles tend to be lighter
 than color-singlet sparticles.
On the other hand, the case $d > 0$ examined in Ref.~\cite{dAMSB2}
 leads to the mass spectrum similar to the GMSB scenario.
In the following, we consider the case $ d > 0$
 based on the simple model discussed in Section~\ref{sec3}.

\section{Minimal model \label{sec5}}
From the above discussion, it is clear that to solve the tachyonic
slepton problem, we need messenger fields which are non-singlet
 under $SU(2)_L\times U(1)_Y$. If we now look at the way type II seesaw
formula for small neutrino mass is implemented \cite{type2}, we find
that we need a pair of $SU(2)_L$ triplet fields,
$\overline{\Delta}:({\bf 3},-1)$ and $\Delta:({\bf 3},+1)$, which
can play the dual role of both generators of neutrino masses as well
as messenger fields.

To see their role in the neutrino sector, we add to the MSSM
superpotential the following couplings of the triplets to the lepton
doublets ($L_i$) and the up-type Higgs doublet ($H_u$) \bea W_{\rm
seesaw} = Y_{ij} L_i \Delta L_j + \lambda H_u \overline{\Delta} H_u,
\eea
 where $i,j$ denotes the generation index,
 and $Y_{ij}$ is Yukawa coupling.
 If they couple to the singlet field
$S$ discussed above as:
\bea
 W_{\rm mess}= S \;  {\rm tr} \left[ \overline{\Delta} \Delta \right].
\label{triplet-mess}
\eea
 then once $\left<S
 \right>\neq 0$, it will give heavy
mass to the triplets.
Integrating out the heavy messengers with mass
 $M_{\rm mess} =\langle S \rangle$,
 this superpotential leads to light neutrino mass matrix
 $M_\nu \sim Y_{ij} \lambda\langle H_u \rangle^2/M_{\rm mess}$.
This is the type II seesaw mechanism. If the messenger
scale lies around the intermediate scale
 $M_{\rm mess} =10^{13-14}$ GeV,
 the seesaw mechanism provides the correct scale
 for light neutrino masses with $Y_{ij}\lambda$ of order one.

Note that since $F_S\neq 0$, the triplets can also serve as
messenger superfields as in usual GMSB models and make additional
contributions to slepton masses. In this minimal case, with a given
$d$ and the formulas in Eq.~(\ref{gauginomass})-(\ref{xi}),
 we now calculate the sparticle mass spectrum including the
 effects of AMSB and anomaly deflection. The beta function parameters
 needed for this purpose are:
 $(b_1, b_2, b_3)=(-33/5, -1, +3)$,
 $(N_1, N_2, N_3)=(18/5, 4, 0)$.
Neglecting the effects of Yukawa couplings \footnote{In general,
there are Yukawa mediation contributions to the $SU(2)_L$ doublet
slepton mass due to the coupling $Y_{ij}L_iL_j\Delta$. In this
paper, we consider the case in which $Y_{ij}\leq 0.1$ by adjusting
the seesaw scale and also parameter $\lambda$, so that the Yukawa
mediation contributions are negligible.},
 the sparticle masses (in GeV) evaluated at $\mu=500$ GeV
 are depicted in Fig.~1 as a function of the messenger scale
 $\log_{10}(M_{\rm mess}/{\rm GeV})$.
Here, we have taken $d=0.81$, $F_\phi=25$ TeV, and
 the standard model gauge coupling constants at the Z-pole as
 $ \alpha_1(m_Z) = 0.0168$, $\alpha_2(m_Z) = 0.0335$
 and $\alpha_3(m_Z) = 0. 118$. Since
 the Higgs triplet pair do not carry color quantum number,
 the gluino mass still stays on the AMSB trajectory and
 does not depend on the messenger scale as showed in the Fig.~1.
Note that for the messenger scale $M_{\rm mess} \gtrsim 10^{14}$ GeV,
 the bino becomes the lightest super particle (LSP) and
 the bino like neutralino would be the candidate of
 the dark matter in our scenario \cite{dAMSB-DM}.
For a small $\tan\beta$, annihilation processes of bino
 like neutralinos are dominated by p-wave and since this annihilation
process is not so efficient, the resultant relic density tends
 to exceed the upper bound on the observed dark matter density.
This problem can be avoided, if the neutralino is quasi-degenerate
 with the next LSP slepton and the co-annihilation process
 between the LSP neutralino and the next LSP slepton can lead to the
right dark matter density.
It is very interesting that our results show
 this degeneracy happening at $M_{\rm mess} \simeq 10^{14}$ GeV,
 which is, in fact, the correct seesaw scale.

In the simple superpotential of singlet discussed in Section~\ref{sec3},
 the messenger scale is given by $M_{\rm mess}=\left<S\right>\sim
\sqrt{F_\phi M}$.
To obtain $M_{\rm mess} \sim 10^{13}-10^{14}$ GeV
 with $F_\phi={\cal O}(10)$ TeV, we can specify the superpotential
 in Eq.~(\ref{superpotential}) as
\bea
 W \sim - m S^2 + \eta \frac{S^4}{M_{Pl}}
\eea with $\eta\sim 10^{-3}-10^{-5}$, where $M_{Pl}$ is the Planck
scale.

\section{Minimal model with grand unification \label{sec6}}
The messengers we have introduced  in the minimal model
 are $SU(3)_c$ singlets, and the existence of such particles
 below the grand unification scale $M_{\rm GUT} \sim 10^{16}$ GeV
 spoils the successful gauge coupling unification in MSSM.
As is well-known, the gauge coupling unification can be kept
 if the messenger fields introduced are in the $SU(5)$ GUT
multiplets. There are two possibilities for such messengers
 that play two different roles in the neutrino sector by the seesaw
mechanism. One is to introduce the messengers of
 ${\bf {15}+\overline{15}}$ multiplets under $SU(5)$, which include
$\Delta$ and $\overline{\Delta}$ as submultiplets. The other
possibility is to introduce ${\bf 24}$ multiplets \cite{typeIII}.

Let us first consider the $\bf 15$ and $\bf\overline {15}$ case
 in the $SU(5)$ GUT model.
We introduce the superpotentials,
\bea
 W_{\rm mess} &=& S \overline{T} T,   \nonumber \\
 W_{\rm seesaw} &=& Y_{ij} \; {\bf \overline{5}}_i \; {\bf \overline{5}}_j \; T
   + \lambda{\bf 5}_H {\bf 5}_H  \overline{T} ,
\eea where $T$ and $\overline T$ are $\bf 15$ and
${\bf\overline{15}}$ multiplets. After integrating the heavy
messengers out, we obtain
 the light neutrino mass matrix as
 $M_\nu \sim \langle {\bf 5}_H \rangle^2 /\langle S \rangle$
 through the type II seesaw mechanism.

Sparticle masses can be evaluated in the same manner as before,
 but in this case, $N_1=N_2=N_3=7$.
The resultant sparticle masses at $\mu=500$ GeV
 are depicted in Fig.~2 as a function of
 the messenger scale $\log_{10}[S/{\rm GeV}]$.
Here, we have taken $d=0.48$ and $F_\phi=25$ TeV. The bino becomes
the LSP, degenerating with
 right-handed sleptons for the messenger scale
 $M_{\rm mess} \sim 10^{13}$ GeV.

In the case of ${\bf 24}$ multiplets ($\Sigma$), the relevant
superpotential is given by \bea
 W_{\rm mess} &=& S \; {\rm tr} [\Sigma^2],  \nonumber \\
 W_{\rm seesaw} &=& Y_i \;  {\bf \overline{5}}_i \; \Sigma \; {\bf 5}_H .
\eea After integrating out the heavy $\bf 24$,
 the light neutrino mass matrix is given by
 $M_\nu \sim Y_i Y_j \langle {\bf 5}_H \rangle^2/\langle S \rangle$.
Note that the rank of this matrix is one.
We need to introduce  at least two ${\bf 24}$ messengers
 to incorporate the realistic neutrino mass matrix.
As an example, we consider two ${\bf 24}$ messengers
 with the same masses.
We evaluate sparticle masses with $N_1=N_2=N_3= 2 \times 5=10$
 in this case.
The resultant sparticle masses at $\mu=500$ GeV
 are depicted in Fig.~3 as a function of
 the messenger scale $\log_{10}[S/{\rm GeV}]$.
Here, we have taken $d=0.35$ and $F_\phi=25$ TeV. The bino becomes
the LSP, degenerate with right-handed sleptons
 for the messenger scale $M_{\rm mess} \sim 10^{13}$ GeV.

\section{Conclusion \label{sec7}}
In conclusion, we have pointed out that a minimal extension of MSSM
needed to explain small neutrino masses via the seesaw mechanism can also
cure the tachyonic slepton mass problem of anomaly mediated supersymmetry
breaking. We have presented the sparticle spectrum for these models and
shown that they can preserve the unification of gauge couplings. We find
it interesting that the same mechanism that explains the smallness of
neutrino masses also cures the tachyonic slepton problem of AMSB.

\acknowledgments
The work of R.N.M. is supported by the National Science Foundation
Grant No. PHY-0652363. The work of N.O. is supported in part by the
Grant-in-Aid for Scientific Research from the Ministry of Education,
 Science and Culture of Japan (\#18740170).
The work of H.B.Y. is supported by the National Science Foundation
 under Grant No. PHY-0709742.

\begin{appendix}
\section{Lifetime of the local minimum }
The scalar potential in Section III $V(S)$ has the global SUSY
minimum
 at the origin, and the minimum we have discussed is a local minimum.
If our world is trapped in the local minimum, it will eventually
 decay into the SUSY minimum.
The life time of the local minimum should be sufficiently long,
 at least, longer than the age of the universe,
 $\tau_U \sim 4.3 \times 10^{17}$ s for our model to be viable.
Here we estimate the decay rate of the false vacuum within the
parameters of our model.

In our calculation, the scalar potential is treated
 in the triangle approximation \cite{DuJe}.
A schematic picture of the scalar potential
 is depicted in Fig.~4.
Let us take the path in the direction of $\Re[S]$:
 climbing up from the local minimum at $\Re[S] = \sqrt{x_+}$
 to the local maximum at $\Re[S] = \sqrt{x_-}$,
 then rolling down to the SUSY minimum at $S=0$.
In the triangle approximation, parameters characterizing
 the potential are
\begin{eqnarray}
  \Delta V_{\pm}, \quad \Delta \Phi_{\pm},
\end{eqnarray}
where $\Delta V_{\pm}$ and $\Delta \Phi_{\pm}$
 are the difference of potential height and
 the distances between the local and global minima
 and potential barrier.
Following Ref.~\cite{DuJe}, we define
\begin{eqnarray}
  c \equiv
  \frac{\Delta V_{-} \Delta \Phi_{+}}{\Delta V_{+} \Delta \Phi_{-}}
\end{eqnarray}
and the decay rate per unit volume is estimated as
 $\Gamma/V \sim e^{-B}$  with
\begin{eqnarray}
 B = \frac{32 \pi^2}{3} \frac{1 + c}{(\sqrt{1 + c} - 1)^4}
   \frac{\Delta \Phi_{+}^4}{\Delta V_{+}} .
\end{eqnarray}
The consistency condition to apply the triangle approximation
 is given by \cite{DuJe}
\begin{eqnarray}
 \left( \frac{\Delta V_{-}}{\Delta V_{+}} \right)^{\frac{1}{2}}
 \geq \frac{2 \Delta \Phi_{-}}{\Delta \Phi_{-} - \Delta \Phi_{+}}.
\label{condition}
\end{eqnarray}

For the scalar potential analyzed in Section~\ref{sec3},
\bea
&& \Delta \Phi_{+} = \sqrt{x_+}-\sqrt{x_-}, \; \;
\Delta \Phi_{-} = \sqrt{x_-} ,  \nonumber \\
&& \Delta V_{+} = V(x_-,0)- V(x_+,0), \; \; \Delta V_{-} = V(x_-,0)
. \eea In order to get the deflection parameter as large as
possible,
 let us consider the case that the local minimum and maximum points
 are very close, namely,
 $\Delta \Phi_{+}$ and $\Delta V_{+}$ are very small.
In this case, the condition Eq.~(\ref{condition}) is satisfied,
 and we can apply the triangle  approximation.
With a small parameter $0 < \epsilon \ll 1$,
 we parameterize
\bea
  m = \frac{5+2 \sqrt{6}}{4} F_\phi \left( 1+\epsilon \right ).
\eea In the limit $\epsilon \to 0$, the local minimum and maximum
 collide and the local minimum disappears.
The deflection parameter is approximately described as \bea
  d \simeq d_{\rm max}-\frac{\sqrt{12+5 \sqrt{6}}}{3} \epsilon^{\frac{1}{2}}
  \simeq d_{\rm max} -1.64 \; \epsilon^{\frac{1}{2}} .
\eea The straightforward calculations give the following results:
\bea
 \Delta \Phi_{+} &\simeq &
  \sqrt{\frac{12 +5 \sqrt{6}}{54+24 \sqrt{6}}}
  \sqrt{F_\phi M} \epsilon^{\frac{1}{2}},  \nonumber \\
 \Delta \Phi_{-} &\simeq &
  \frac{1}{2} \sqrt{\frac{9 + 4 \sqrt{6}}{6}}
  \sqrt{F_\phi M},  \nonumber \\
\Delta V_{+} &=& \frac{(12+5 \sqrt{6})^{\frac{3}{2}}}{27}
  F_\phi^3 M \epsilon^{\frac{3}{2}},   \nonumber \\
\Delta V_{-} &=&
  \frac{1107 + 452 \sqrt{6}}{288}
  F_\phi^3 M .
\eea Also, we find \bea B \simeq
 \frac{\pi^2  128 (12+5 \sqrt{6})^{\frac{3}{2}}}
{9 (6937+2832 \sqrt{6})}
  \frac{M}{F_\phi}  \epsilon^{\frac{3}{2}}
\simeq 1.21 \times   \frac{M}{F_\phi}  \epsilon^{\frac{3}{2}} .
\eea Recalling that the messenger scale is roughly given by
 $M_{\rm mess} \sim \sqrt{F_\phi M}$
 and $ F_\phi \simeq 10$ TeV to obtain sparticle masses
 around 100 GeV - 1 TeV, we can rewrite $B$ as
\bea
 B \simeq
 1.21 \left(\frac{M_{\rm mess}}{F_\phi} \right)^2 \epsilon^{\frac{3}{2}}
 =
 1.21 \times 10^{20}
 \left(\frac{M_{\rm mess}/10^{14}\; {\rm GeV}}{F_\phi/10\; {\rm TeV}}
 \right)^2 \epsilon^{\frac{3}{2}} .
\eea For the parameters chosen in Fig.~1,
 $M_{\rm mess} \simeq 10^{14}$ GeV,
 $F_\phi=25$ TeV,
 $d=0.81$ corresponding $\epsilon \simeq 1.57 \times 10^{-5}$,
 we find $B \simeq 1.20 \times 10^{12}$.
The life time of the local minimum is extremely long.
\end{appendix}


\newpage

\begin{figure}[t]
\includegraphics[scale=1.2]{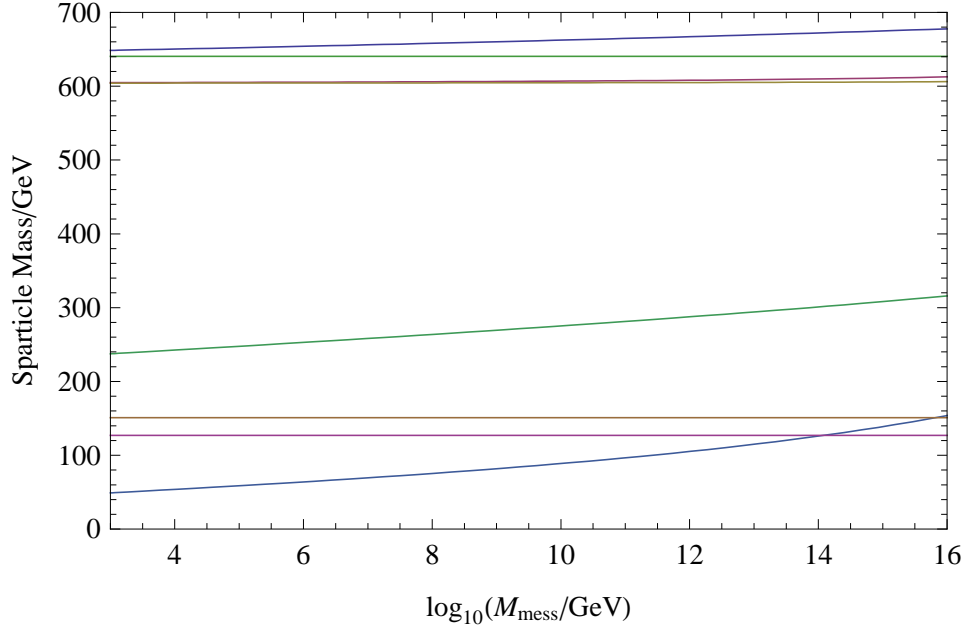}
\caption{
Sparticle masses at $\mu=500$ GeV
 as a function of at the messenger scale
 in the type II seesaw model with one pair of
 $SU(2)_L$ triplet messengers.
Here $d=0.81$ and $F_\phi=25$ TeV have been taken.
Each line corresponds to
 the left-handed squark ($m_{\tilde{Q}} $),
 the gluino ($M_3$),
 the right-handed up-squark ($m_{\tilde{u^c}}$),
 the right-handed down-squark ($m_{\tilde{d^c}}$),
 the left-handed slepton ($m_{\tilde{L}}$),
 the Wino ($M_2$), the bino ($|M_1|$), and
 the right-handed slepton ($m_{\tilde{e^c}}$)
 from above at $M_{\rm mess}=10^3$ GeV.
Two lines of $m_{\tilde{u^c}}$ and $m_{\tilde{d^c}}$
 are overlapping and not distinguishable.
For the messenger scale $M_{\rm mess} \gtrsim 10^{14}$ GeV,
 the bino becomes the lightest super particle.
}
\end{figure}
\begin{figure}[t]
\includegraphics[scale=1.2]{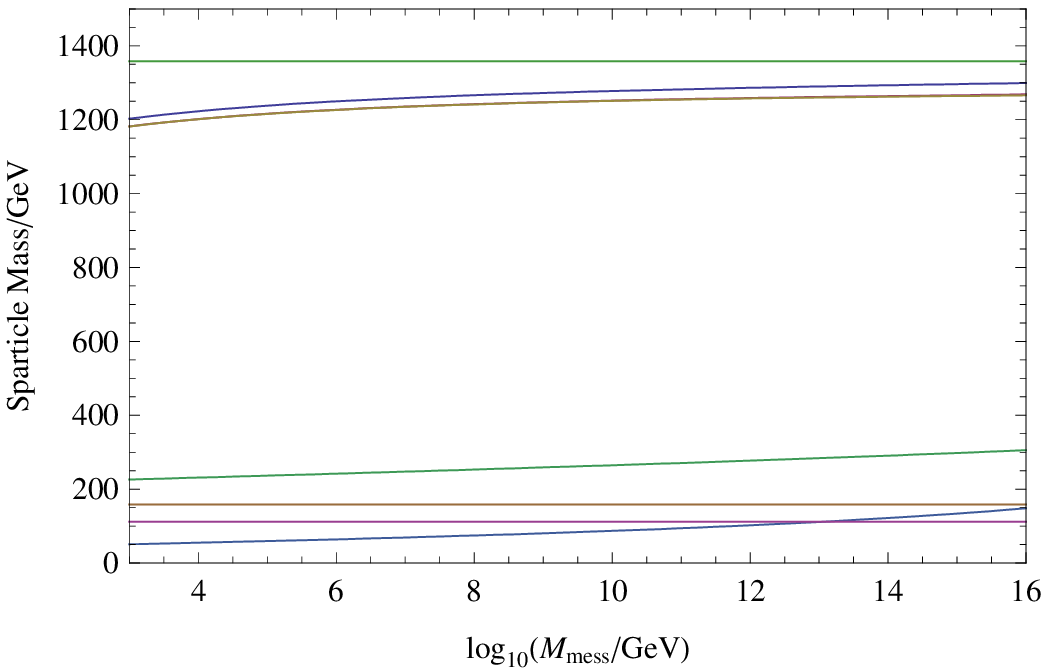}
\caption{
Sparticle masses at $\mu=500$ GeV
 as a function of at the messenger scale
 in the type II seesaw model with one pair of
 ${\bf \overline{15}}+{\bf 15}$ messengers.
Here $d=0.48$ and $F_\phi=25$ TeV have been taken.
Each line corresponds to
$M_3$, $m_{\tilde{Q}} $, $m_{\tilde{u^c}}$, $m_{\tilde{d^c}}$,
$m_{\tilde{L}}$, $M_2$, $|M_1|$, and $m_{\tilde{e^c}}$
 from above at $M_{\rm mess}=10^3$ GeV.
Two lines of $m_{\tilde{u^c}}$ and $m_{\tilde{d^c}}$
 are overlapping and not distinguishable.
For the messenger scale $M_{\rm mess} \gtrsim 10^{13}$ GeV,
 the bino becomes the LSP.
}
\end{figure}
\begin{figure}[t]
\includegraphics[scale=1.2]{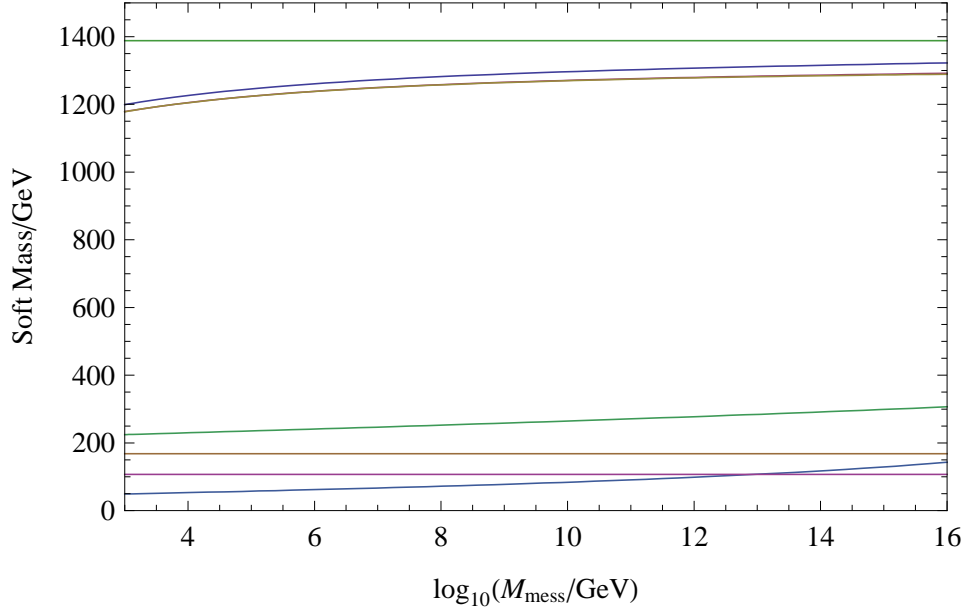}
\caption{
Sparticle masses at $\mu=500$ GeV
 as a function of at the messenger scale
 in the model with two pairs of ${\bf 24}$ messengers.
Here $d=0.35$ and $F_\phi=25$ TeV have been taken.
Each line corresponds to
 $M_3$, $m_{\tilde{Q}} $, $m_{\tilde{u^c}}$, $m_{\tilde{d^c}}$,
 $m_{\tilde{L}}$, $M_2$, $|M_1|$, and $m_{\tilde{e^c}}$
from above at $M_{\rm mess}=10^3$ GeV.
Two lines of $m_{\tilde{u^c}}$ and $m_{\tilde{d^c}}$
 are overlapping and not distinguishable.
For the messenger scale $M_{\rm mess} \gtrsim 10^{13}$ GeV,
 the bino becomes the LSP.
}
\end{figure}
\begin{figure}[t]
\includegraphics[scale=0.6]{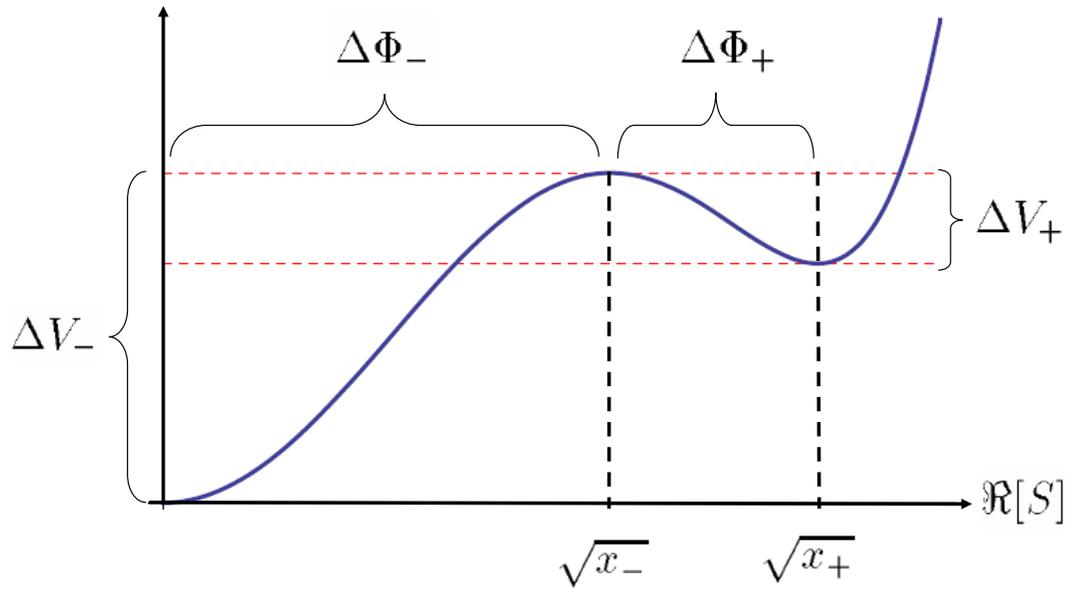}
\caption{
Schematic picture of the scalar potential $V(S)$
 as a function of the real part of $S$.
}
\end{figure}

\end{document}